 \documentstyle[aps]{revtex}
\begin{document}
\draft
\title{
$\Gamma$ phonons and microscopic structure of orthorhombic KNbO$_3$ \\
from first-principles calculations
}
\author{A.~V.~Postnikov,\cite{*} and G.~Borstel}
\address{
Universit\"at Osnabr\"uck -- Fachbereich Physik,
D-49069 Osnabr\"uck, Germany}
\date{\today}
\maketitle
\begin{abstract}
{}From a series of total energy calculations
by the full-potential linear muffin-tin orbital method,
the total energy hypersurface as function of
atomic displacements from equilibrium positions
has been fitted for different $\Gamma$ phonon modes
in orthorhombic KNbO$_3$.
Frequencies and eigenvectors of all TO $\Gamma$
phonons have been calculated in the harmonic
approximation, and in the quantum oscillator
scheme -- for $A_2$ and $B_2$ modes. The microscopic
structure of the orthorhombic phase has been
analyzed in a series of supercell calculations
for different patterns of Nb displacements,
providing indications in favour of the chain
structure, with oppositely directed neighboring
chains.
\end{abstract}

\section{Introduction}
\label{sec:intro}

Of three ferroelectric phases known for potassium niobate,
the orthorhombic phase is probably
the best experimentally studied. This is a room-temperature
phase which exists in the temperature range $-$10 to 225 $^0$C
and is therefore the most important for any practical
applications, e.g., related to the photorefractive properties
of KNbO$_3$.
Considerable amount of information has been accumulated
from Raman scattering \cite{bh76,qbkr76,frkc88} and neutron scattering
\cite{ccdw74} experiments; local environment of Nb atoms has been
probed by EXAFS measurements \cite{mphd93};
recent experiments on impulsive simulated (time-resolved)
Raman scattering \cite{dwng92,tpdoc} provide additional data on
oscillatatory and relaxation-type phonon modes.
However, the understanding, on the theoretical level,
of fine mechanisms behind ferroelectric atomic displacements
and phase transitions in KNbO$_3$ is still not sufficient.
{\em Ab initio} calculations which may clarify this point
have not been up to now performed, to our best knowledge, for the
orthorhombic phase, because of the lower symmetry
and correspondingly higher calculational effort demanded,
as compared to the cubic (non-polar) phase
and the tetragonal ferroelectric phase.

In the present paper, we extend our previous research
on the equilibrium geometry of the ferroelectric ground
state structure of KNbO$_3$ \cite{ktn3} and on phonon
properties in cubic and tetragonal phases \cite{phonon}
onto the orthorhombic phase of this material.
The band structure and total energy calculations,
based on which the equilibrium geometry and phonon
properties are investigated, have been performed
using the full-potential linear muffin-tin orbital code
by Methfessel \cite{msm1,msm2}. The technical aspects
of the calculations for KNbO$_3$ are discussed in some detail
in Ref. \cite{ktn3}.

Apart from lower symmetry, another difficulty in performing
the calculations for the orthorhombic phase is the much smaller
magnitude of relevant energy differences, e.g., related
to the displacements within the soft mode, as compared
to cubic or tetragonal phases. This demands
an accuracy of total energy calculation of at least 0.01 mRy.
In the computer code we applied,
that is technically possible
due to the use of the Harris energy which coincides with the
Kohn--Sham energy in the first order with respect to variations
around the self-consistent density and converges much more
smoothly than the latter. (More details on the properties
of the Harris functional and its applications may be found
in Ref. \cite{forces}). Also, the convergence of the
total energy in the number of special points in
the Brillouin zone has to be particularly strictly controlled.
The corresponding error was always
within 0.01~mRy in our calculations
even for different supercell geometries (with different
patterns of the {\bf k}-points for integration) if they
actually refer to the same crystal (super)structure \cite{note1}.

The paper is organized as follows: In section \ref{sec:struc},
the geometry optimization in the orthorhombic phase is discussed
and compared with the experimental data. In section \ref{sec:phonons},
the results of the frozen phonon calculations in the harmonic
approximation are presented, and in section \ref{sec:quant} --
the analysis of phonon frequencies in terms of the quantum
oscillator model. Finally, in section \ref{sec:finite}
different models for the microscopic structure of the orthorhombic
phase are considered, and the relevance of the chain
structures is emphasized.

\section{Optimized structure of the orthorhombic phase}
\label{sec:struc}

In discussing the geometry of the orthorhombic phase,
we follow the notations introduced by Hewat \cite{hewat}:
three lattice vectors are $a$=3.973 {\AA} along
${\bf x}=[100]$ of the cubic aristotype;
$b$=5.695 {\AA} along ${\bf y}=[0\bar{1}1]$ and
$c$=5.721 {\AA} along ${\bf z}=[011]$.
The positions of atoms and the magnitudes of
their off-center displacements in terms of
these lattice vectors are given in Table \ref{tab:equilib}.
It would be in principle possible in a first-principle
calculation to minimize the total energy
with respect to all seven independent structure parameters,
i.e., three lattice vectors and four internal
atomic displacements. The optimization of such
multidimensional function is however computationally
very demanding and hardly likely to produce any new
essential information. Based on our previous experience,
gained from the study of tetragonal KNbO$_3$, that
the lattice strain found from the total energy minimization
is in very good agreement with experiment \cite{ktn3},
and in order to keep the number of independent parameters
in the structure optimization manageable, we fixed
the lattice volume and strain in our
frozen phonon calculations and used for the lattice
parameters the experimental data of Ref.\cite{hewat}.
However, all off-center displacements of atoms,
which are allowed by symmetry, have to be optimized
in the calculation, in order to guarantee that for each phonon,
the force constants are calculated near the total energy
minimum. The optimization has been performed by
the polynomial fit of the total energy as
function of four variables; the values of parameters
corresponding to the total energy minimum are also given
in Table \ref{tab:equilib}.
Similarly to the choice of Ref.\cite{hewat}, we kept Nb
in the center of the cube and allowed all
other atoms to relax with respect to it.
Comparing the calculated and experimental values of
$\Delta$'s in Table \ref{tab:equilib}, one can conclude
that the $z$-displacement of oxygen atoms is fairly well
reproduced, including the fact that O$_1$ is
slightly farther shifted than O$_2$, O$_3$.
The $y$-displacement of O$_2$ is larger than the experimental
estimate; it should be noted however that the magnitude
of this displacement is very small, and of primary importance
is the fact that the calculation correctly reproduces
the sign of this parameter.
Finally, the $z$-displacement of K is overestimated
by almost a factor of 3, that is identical
to what we have found earlier when optimizing the geometry
of the tetragonal phase \cite{phonon}. In other words,
in our calculations K tends to remain
coupled to the oxygen sublattice,
so that the ferroelectric distortion arises roughly
due to the pure displacement of Nb.

\section{Frozen phonons}
\label{sec:phonons}

Combined atomic displacements in the orthorhombic phase which has
the space group $Amm2$ (No. 38 of the International tables)
are split by symmetry into four one-dimensional
irreducible representations. After projecting out
three translational modes similarly to how it was
discussed in some
detail by Cohen and Krakauer for cubic BaTiO$_3$ \cite{cohen},
one stays with 12 symmetry coordinates $S_t$. The choice
of $S_t$ (arbitrary up to constructing other
linear combinations within each irreducible representation)
we used in our total energy fitting and frozen phonon
calculations, in terms of cartesian displacements of atoms
1 to 5 according to Table \ref{tab:equilib}, is given
in Table \ref{tab:symmet}.

We calculated the total energy within each irreducible
representation for a sufficient number of different
displacement patterns as to make a fit by the fourth-order
polynomial. Apart from extracting second derivatives
for the force constants, this fit made possible the
quantum oscillator analysis described below in section
\ref{sec:quant} and the optimization of finite
displacements within the soft mode. A list of polynomial
coefficients for the $B_2$ representation, that is
probably of the most general interest since it contains
the soft mode, is given in Table \ref{tab:polynom}.
Total energy in mRy may be restored as
the sum over all coefficients multiplied by
corresponding powers of $S_t$ in a.u.

In constructing the kinetic-energy matrix and
solving the eigenvalue problem for the TO $\Gamma$ phonons
in the harmonic approximation, we followed the guidelines
of Ref. \cite{wdc55} (see also Ref. \cite{cohen} for
some details). Calculated frequencies are given
in Table \ref{calc:knoort} along with the eigenvectors,
which are expressed in terms of cartesian displacements
of individual atoms (relative to the center of mass),
multiplied by the square roots of masses, via back transformation
from the symmetry coordinates defined in Table \ref{tab:symmet}.
The comparison with experimentally measured frequencies
is given in Table \ref{exp:knoort}.
The overall agreement with experiment is within about 10\%,
with the exception of TO$_2$ $B_1$ mode and the $A_2$ mode.
For the $B_1$ phonon, the difference is probably due to
some inaccuracy in the polynomial fit to
the four-dimensional total energy hypersurface,
since the frequency of the corresponding vibration
(K -- Nb stretching on the background of relatively undistorted oxygen
octahedra) in the tetragonal phase (where it belongs to the
$A_1$ mode) came out in much better agreement with experiment
in our previous calculations of Ref. \cite{phonon}. For the $A_2$
phonon, the difference seems to be due to anharmonicity effects
as has been already supposed in Ref. \cite{phonon}.
The explicit form of the
total energy fit for this mode, which we found to be
$\Delta E$ (mRy) = 15.16 $(S_5)^2$ + 12.99 $(S_5)^4$
($S_5$ in a.u.), makes it possible to calculate the
relevant frequency beyond the harmonic approximation, that is done
in Section \ref{sec:quant}.

\section{Quantum oscillator treatment}
\label{sec:quant}

Since the total energy as function of symmetrized
displacements was available from our calculations
in a sufficient number of points for providing
the fourth-order polynomial fit, an extension of
our analyzis beyond the harmonic approximation
becomes feasible. Rather than introducing anharmonic
corrections to frequencies obtained in the previous
section, we prefered to solve the Schr\"odinger equation
for an, in principle, multidimensional quantum oscillator
with a general-shape potential. We solved the Schr\"odinger
equation numerically, by a finite-difference method
(see, e.g., Ref.\cite{mitch,ames}), so that no basis
functions have been employed.

Similarly to how it was described elsewhere for
the classical harmonic oscillator \cite{cohen},
we use an arbitrary convenient set of symmetry-adapted
displacement coordinates
$S_t=\sum_i B_{ti}x_i$
($x_i$ are conventional cartesian displacements),
which form a complete basis within a particular
irreducible representation, but do not need to be
orthonormal. Then, the Schr\"odinger equation
aquires a form:
\[
 \Big[- \frac{~\hbar^2}{2}\sum_{tt'}
     \frac{\partial}{\partial S_t}G_{tt'}
     \frac{\partial}{\partial S_{t'}} +
     V(\{S_t\})\Big] \Psi = E \Psi
\]
where $G_{tt'}=\sum_i B_{ti} m^{-1}_i B_{t'i}$
is the kinetic-energy matrix.
Mixed derivatives can further be excluded
(that is desirable for maintaining high
sparseness of the finite-difference matrix)
by the following orthogonalizing transformation:
\[
 Q_t = \sum_{t'} \frac{X_{t't}}{\sqrt{\lambda^t}}\,S_{t'},
\]
where $X_{t't}$ is the $t$-th eigenvector, corresponding
to the eigenvalue $\lambda^t$, of the kinetic energy
matrix.

The energy differences between the zero-point energy level
and subsequent higher states give rise to characteristic
frequencies of several lowest phonon modes. For the
total energy fitted by the second order polynomial,
these lowest frequencies should exactly coincide
with the classical frequencies obtained in the harmonic
approximation. This provides an additional mean of
testing the accuracy of the finite-difference calculation
scheme. Apart from allowing the anharmonicity in the
most general way, the quantum treatment provides
the description of high-frequency phonon branches
(i.e. those lying beyond the lower $3N-6$ optical modes)
and paves a way to incorporating the temperature effects.
The energy levels of the oscillator may be
further populated
according to the Bose--Einstein statistics,
and the transitions between levels analyzed.
An analysis of this kind has been done by Bakker {\em et al.}
\cite{bakker} for a one-dimension quantum oscillator model
of the $A_1$ phonon in LiTaO$_3$.

The accuracy of the
calculation is essentially limited by the finiteness
of the discrete mesh, and the convergence of
eigenvalues in this parameter was explicitly controlled;
however for the 4-dimensional oscillator ($A_1$ and B$_2$
irreducible representations), the maximal manageable grid
of $14^4$=38416 points was not sufficiently dense
to produce as stable frequencies as those for $B_1$
and $A_2$ modes.

For the $B_2$ mode, the lowest energy levels of the
3-dimensional quantum oscillator were found to be
64, 89, and 130 cm$^{-1}$. Thus, the lowest possible
phonon frequency is 25 cm$^{-1}$, that is
of correct order of magnitude, as compared with the
experimental soft mode frequency (the harmonic
approximation could only provide an information that
the corresponding mode is indeed soft). Because of the
small energy differences between several lowest
oscillator levels, corresponding transitions
can be easily excited already at medium temperatures,
thus leading to a substantial increase of the
soft mode frequency on heating. We however did not
consider this effect in detail.

For the one-dimensional $A_2$ problem, the five lowest energy
levels of the quantum oscillator are: 116, 357, 612, 880
and 1160 cm$^{-1}$. The lowest transition energy is
therefore 241 cm$^{-1}$ that is already an improvement
over the result obtained in the harmonic approximation.
The gradual increase of the separation between
subsequent higher levels would result in somehow higher
mean frequencies at elevated temperatures, but
this effect is expected to be much smaller than for the
soft mode.

\section{Atomic displacements within the soft mode}
\label{sec:finite}

Now that the whole total energy hypersurface is mapped as function
of symmetrized displacements of atoms from their equilibrium
positions in the orthorhombic structure, it becomes possible
to analyze the finite distortion compatible with the soft $B_2$ mode
which is a precursor of the orthorhombic to rhombohedral
phase transition. We did not consider the change in the
lattice strain which is an important element of such transition.
Nevertheless, it makes some sense to compare the symmetry-lowering
displacements of atoms from equilibrium positions
in the orthorhombic phase with the experimental geometry
of the rhombohedral phase.

We performed some additional fine-mesh fit near the position
of the total energy minimum, in order to determine the
$x$-displacements of atoms with better accuracy than
it is possible from the global fit by the 4th order polynomial.
The equilibrium displacements of K, Nb and O(I) in the
[100] direction from the positions listed in Table \ref{tab:equilib}
are 0.006 {\AA}, $-$0.043 {\AA} and 0.013 {\AA}, correspondingly;
they result in the total energy lowering of 0.06 mRy per formula
unit. An additional degree of freedom enabling the displacement
of Nb in the {\bf z} direction along with that in {\bf x}
(i.e. the relaxation within both $B_2$ and $A_1$ modes)
did not result in any better total energy minimum.
The relative positions of thus displaced atoms
projected on the ({\bf x},{\bf z})
plane of the orthorhombic cell are shown in Fig. \ref{fig:displa}
in comparison with the experimentally determined positions
in the rhombohedral phase.

Magnitudes of the {\bf x}-displacements of Nb and O
in the rhombohedral phase are somehow larger than within the ''frozen''
soft mode. Otherwise,
the most pronounced difference is the underestimated
{\bf z}-displacement of K,
as has been already mentioned in section \ref{sec:struc}.

It is of considerable interest to know whether the displacements
compatible with the $B_2$ symmetry are energetically favourable
if they occur only globally, simultaneously over the crystal, or
also locally, on a short-range scale, as may be expected
from the eight-site model
(see, e.g., Ref. \cite{dwng92} for the discussion on the latter).
We addressed this problem
early for the case of tetragonal soft-mode displacements
from the cubic phase \cite{willi}, and it was established that
the [001] displacements result in a total energy lowering
when they occur simultaneously in all unit cells, but not
if they occur only
in 25\% of cells, as was tested in a supercell calculation.
In other words, it is the indication that the cubic to tetragonal
ferroelectric transition cannot be triggered by a spontaneous
displacement of a single Nb atom, but rather needs a larger
self-supporting region where such transformation occurs. \cite{note2}
Unfortunately, we could not give any estimates as for the size
or shape of such regions from our supercell calculations.
With respect to the orthorhombic phase of KNbO$_3$, the
similar theoretical study of the effect of local displacements
was motivated by recent experimental observation
by Dougherty {\it et al.} \cite{tpdoc} of relaxation
processes in time-resolved Raman spectra
which have been attributed to the hoppings,
compatible with the $A_1$ mode, over three classes of
off-center Nb sites, distinguished by symmetry. There was
however no experimental information available as for
the spatial extent of spatial correlations in such
relaxation processes.

We tried to analyze several patterns of local displacements
in the orthorhombic phase in supercell calculations,
in order to establish some most important trends.
Since it was not feasible to optimize the geometry
in a supercell calculation in terms of all relevant atomic
displacements, we limited ourselves to the study of the effect
of pure Nb displacements which are obviously dominant.
The incorporation of the relaxation of other atoms
may somehow correct the magnitudes of the energy differences,
but hardly the underlying qualitative trends. We considered
local Nb displacements along {\bf z} and {\bf x}, i.e. those
compatible with $A_1$ and $B_2$ modes, from the equilibrium
position defined in Table \ref{tab:equilib}. The relevant
crystal superstructures modelled by 2$\times$KNbO$_3$
and 4$\times$KNbO$_3$ supercells are schematically shown
in Figures \ref{fig:trends1} and \ref{fig:trends2},
along with corresponding total energy dependencies.

Fig. \ref{fig:trends1} shows the effect on the total energy
of {\bf z}-displacements
of every fourth ''impurity'' Nb atom, while other three
Nb's in the supercell are kept fixed and provide a bulk
macroscopic polarization. Apart from the minimum corresponding
to $\Delta_z$(Nb)=0., the total energy grows rapidly;
the displacement against the electrostatic field ($A_1$ mode)
through the centre of the O$_6$ nearest-neighbor octahedra
to the symmetric position increases the total energy
by about 30 mRy. Compared to the tiny value of
0.06 mRy energy lowering due to subsequent {\bf x}-displacement
($B_2$ mode) into one of ''eight sites''
nearest to this point, it leaves
no chance to find there a local minimum of the total energy
with respect to both $\Delta_z$ and $\Delta_x$ variables.
The contradiction with the experimental evidence
of Ref.\cite{tpdoc} that there {\em are} relaxation-type
hoppings of the $A_1$ symmetry between distinct local
minima is apparently resolved by realizing that
in the experiment, a considerable number of neighboring Nb atoms
undergoes a relaxation hopping simultaneously, depending
on the energy pumped in.
A limiting case where really {\em all} atoms in crystal
are involved in this process would mean the inversion
of polarization, correspondingly mapping the total
energy minima into new positions through the center of
the unit cell.

Another choice of 4$\times$KNbO$_3$ supercell
where ''impurity'' Nb atoms
are organized in chains along the {\bf z} direction
(Fig. \ref{fig:trends1}b) leads to the same type
of the energy dependence versus the displacement magnitude.
It indicates that the local displacements against the
bulk polarization are the main energetically unfavourable
factors, whereas local frustrations within the {\bf z}-chains
of interchanging ''impurity'' and ''bulk'' atoms,
which are peculiar for the first supercell geometry,
are of minor importance.

In contrary to this type of behavior, the {\bf x}-displacements
normal to the direction of macroscopic polarisation ($B_2$ mode)
were found to be energetically favourable
even if not supported by a similar displacement of near neighbors.
This is in itself a clear indication of the
order-disorder phase transition type from orthorhombic to
rhombohedral structure. The precise energetics for different patterns
of ''up''- and ''down''- Nb displacements from the ({\bf y},{\bf z})
plane deserves however more detailed investigation. In addition
to the perfect phase with the ''frozen out'' $B_2$ phonon
(Fig. \ref{fig:trends2}a), we considered two types of
2$\times$KNbO$_3$ supercells which combine ''up'' and ''down''
displacements in two different ways: a chessboard arrangement
with chains of identically shifted Nb atoms in the
{\bf x} direction (Fig. \ref{fig:trends2}b) and
a NaCl-type structure where such chains are broken
(Fig. \ref{fig:trends2}c). The corresponding total energy
differences show that the last structure with frustrations
in the {\bf x} direction due to broken chains in absolutely
unstable. Among two other structures which both contain
undistorted chains, the one where ''up'' and ''down''-chains
appear in mixture is definitely more stable.
An important consequence from this fact is its full agreement
with the ''chain structure'' of the orthorhombic phase as proposed
by Comes {\em et al.} \cite{comes}, whereas any other mixture
of ''up''- and ''down'' displacements which incorporates
frustrations in the {\bf x} direction can be relatively
safely discarded. Another important concequence regards the
mutual arrangement of chains. There seems to be a certain
correlation within the ({\bf y},{\bf z}) plane which favours
each chain to have the neighboring chains with the opposite
direction of displacement. In order to check that this
statement is not confined to the particular chessboard
geometry of the supercell in Fig. \ref{fig:trends2}b,
we performed an additional calculation where the effect
of the displacement reversal within one chain out of four
in the initial ''perfect'' geometry of Fig. \ref{fig:trends2}a
has been studied. This was our possibly best supercell
approximation to the displacement reversal within just one
single chain in crystal at all, because thus reversed chains do not
contact in this geometry.
The result is that
the spontaneous, say, ''down'' reversal of Nb displacements within one
chain, with respect to the uniform background of ''up''
displacements, lowers the total energy by $\sim$0.01 mRy per formula
unit, i.e. {\em is} energetically favourable.
The interaction between chains then works until some
particular pattern with equal number of ''up'' and ''down''
chains is established. This stabilizes the orthorhombic
structure, unless the lattice strain is allowed
to become an independent parameter, and the system
finds another global energy minimum compatible
with the rhombohedral space group.

Since the energy lowering due to {\bf x}-displacements
is smaller than the zero-point energy of the quantum
oscillator in a potential well related to the $B_2$ mode
vibrations, it is worth noting that the above mentioned
finite ''up'' and ''down'' displacements should not
be understood literally, like well defined ''frozen'' atomic
positions at zero temperature.
Rather, we have to do with a damped quantum solution
(in agreement with experimental observation
of heavily damped $B_2$ modes, Ref. \cite{tpdoc})
which is symmetric with respect to ''up'' and ''down''
{\bf x}-directions for each single quantum oscillator. The interaction
between chains distorts
the symmetry of each individual solution due to
the coupling between oscillators, and the mean value
of corresponding {\bf x}-displacement gets accordingly
displaced. Its average over the whole system
however remains zero.

\section{Summary}

We extended our previous theoretical study of TO $\Gamma$ phonons
in KNbO$_3$ to the orthorhombic phase. Based on the
total energy calculations and polynomial fit of the results
versus atomic displacements compatible with different
irreducible representations, classical frozen phonon
frequencies and corresponding eigenvectors
have been calculated in the harmonic approximation,
in satisfactory agreement with experimental data.
For $A_2$ and $B_2$ modes, the quantum oscillator analysis
has been done as well, resulting in some corrections
to classical harmonic frequencies. The microscopic
structure of the orthorhombic phase has been further
analyzed in a series of supercell calculations
modelling several patterns of Nb [100]-displacements
compatible with the $B_2$ mode. It was established
that the formation of chains of uniformly displaced
Nb atoms along the [100] direction is highly
energetically favourable, in agreement with the
experimental observations of the chain structure.
The interaction between chains then favours
the zero value of the average [100] Nb displacement,
stabilizing the orthorhombic structure.

\acknowledgements

The authors are grateful to M. Methfessel for his assistance and
advise in using his full-potential LMTO code, and to
T. Dougherty and H. Bakker for providing manuscripts
of their papers prior to publication and for stimulating discussions.
Consultations with by M. Shamonin and P. Hertel were essential for
getting acquaintance of practical finite-difference schemes.
Financial support of the Deutsche Forschungsgemeinschaft
(SFB~225, Graduate College) is gratefully acknowledged.

\begin{figure}
\caption{
Displacements of atoms, relative to the center of mass,
in the ({\bf x},{\bf z}) plane of orthorhombic KNbO$_3$:
from geometry optimization based on total energy
calculations (black circles);
from neutron diffraction measurements in the rhombohedral phase,
Ref. \protect\cite{hewat} (open circles).
}
\label{fig:displa}
\end{figure}

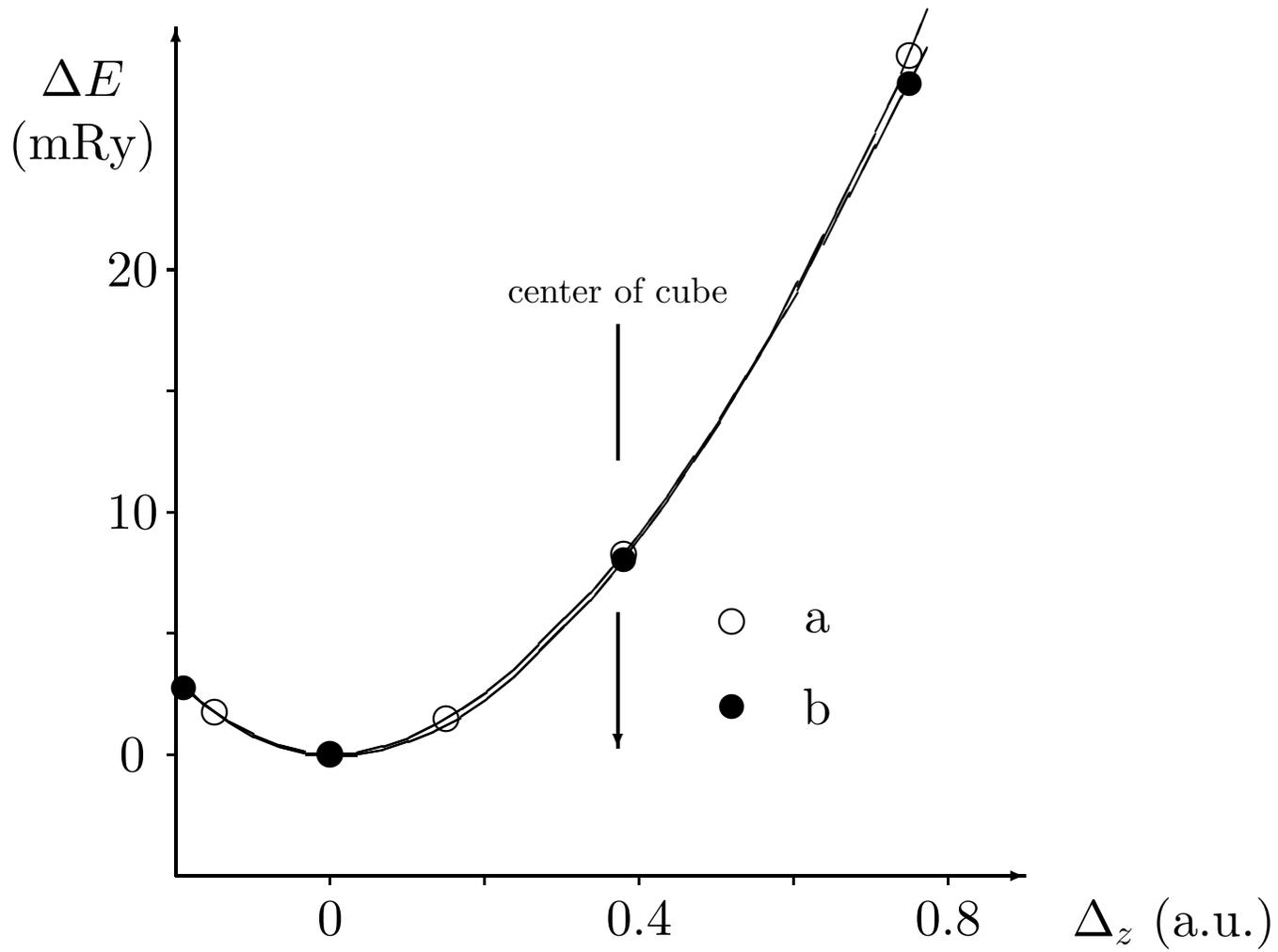
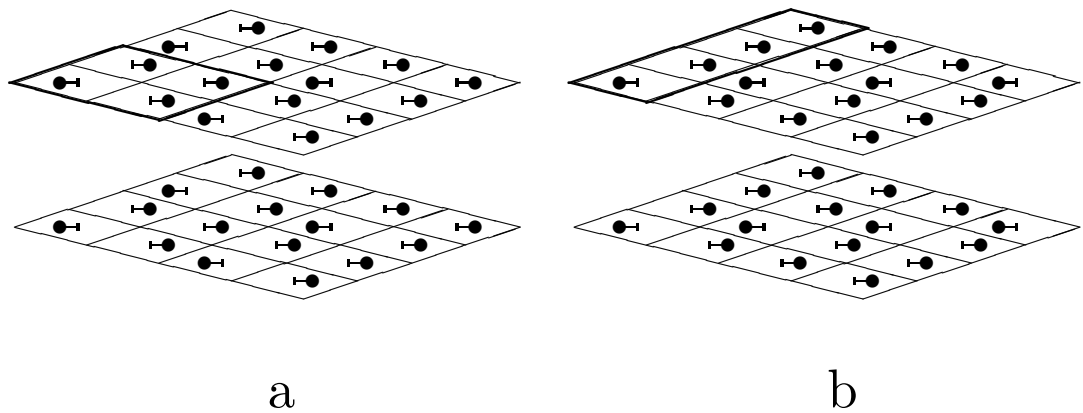
\begin{figure}
\caption{
Total energy differences (per formula unit)
versus {\bf z}-displacements of Nb from
the equilibrium position for two superstructures.
Thick line contours the supercell used.
}
\label{fig:trends1}
\end{figure}

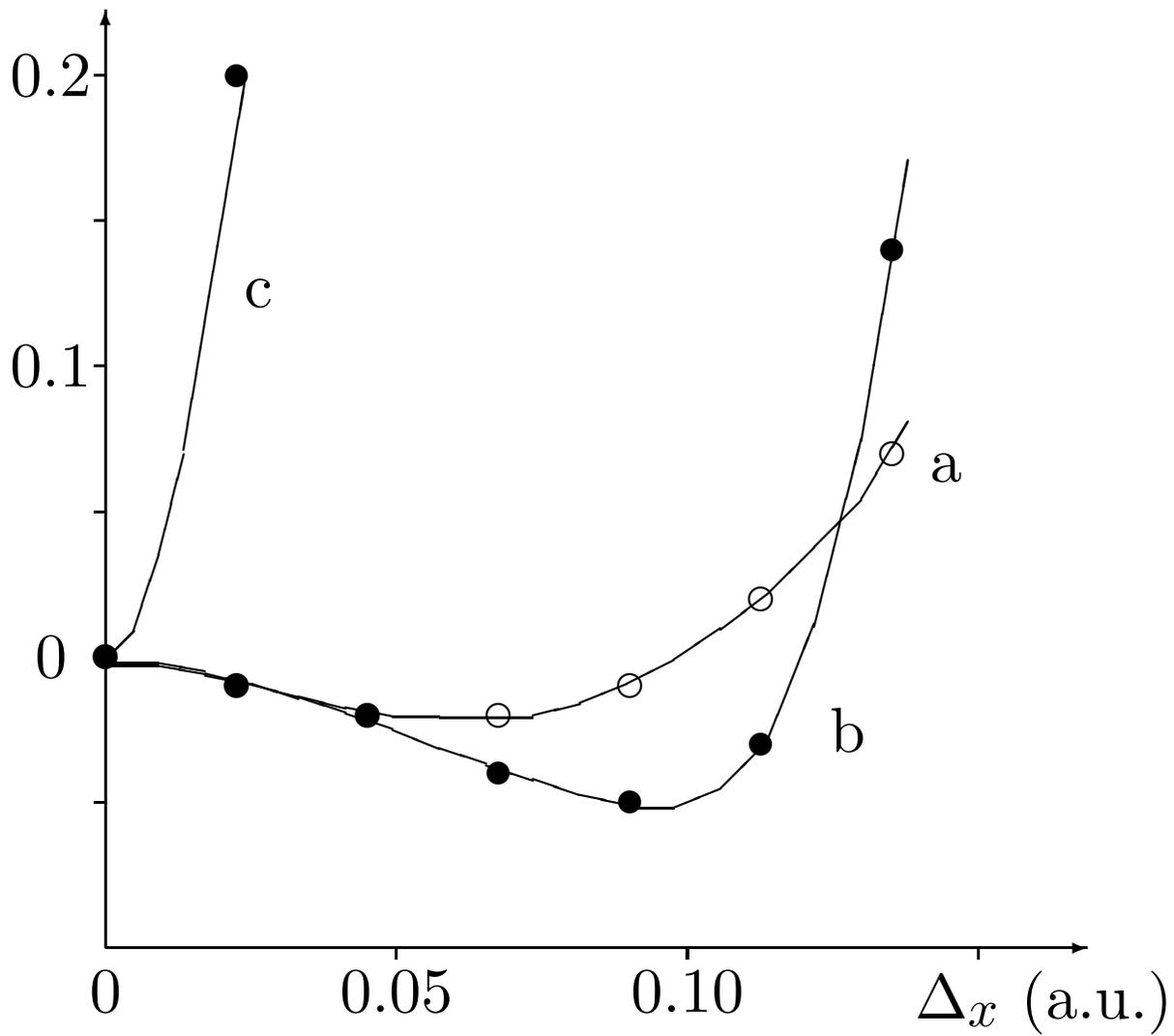
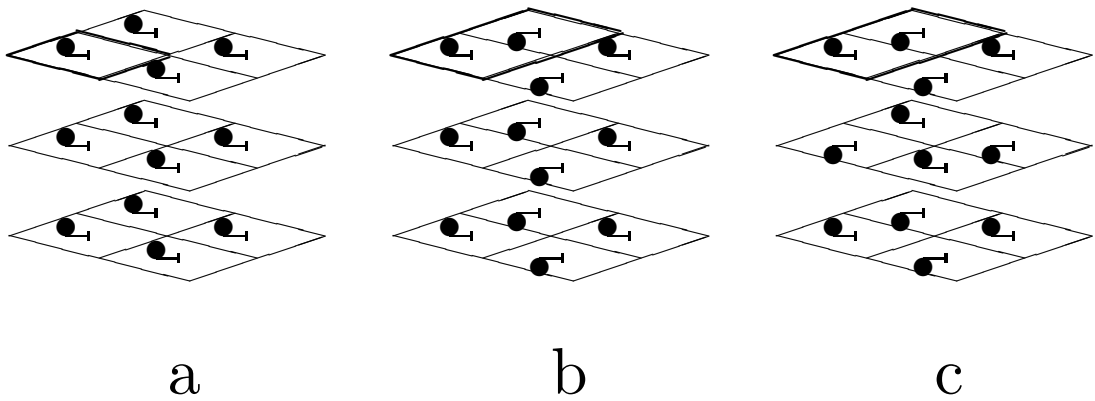
\begin{figure}
\caption{
Total energy differences (per formula unit)
versus {\bf x}-displacements of Nb from
nominally equilibrium position in the orthorhombic
structure for three superstructures.
Thick line contours the supercell used.
}
\label{fig:trends2}
\end{figure}

\begin{table}
\caption{
Positions of atoms in orthorhombic phase of KNbO$_3$
(in terms of lattice parameters) as determined
by neutron diffraction measurements,
Ref. \protect\cite{hewat}, and optimized
in the FP-LMTO total-energy calculation.
}
\label{tab:equilib}
\begin{tabular}{cllllrldd}
   & Atom & $a$ & $b$ & $c$ & & & ~~~~$\Delta_{exp}$ &
                               ~~~~$\Delta_{calc}$ \\
\hline
 1 & K     & 0~~~~~~ & 0 & $\Delta_z$ & & & 0.0138 & 0.0367 \\
 2 & Nb    & $\frac{1}{2}$ & 0 & $\frac{1}{2}$ \\
 3 & O(I)  & 0 & 0 &
             $\frac{1}{2}+\Delta_z$ & & & 0.0364 & 0.0326 \\
 4 & O(II) & $\frac{1}{2}$ & $\frac{1}{4}+\Delta_y$ &
             $\frac{1}{4}+\Delta_z$ &
             & $\Delta_z$~: & 0.0342 & 0.0323 \\
 5 & O(II) & $\frac{1}{2}$ & $\frac{3}{4}-\Delta_y$ &
             $\frac{1}{4}+\Delta_z$ &
 \raisebox{2.5ex}[0pt]{$\Bigg\}$} &
             $\Delta_y$~: & $-$0.0024 & $-$0.0062 \\
\end{tabular}
\end{table}

\begin{table}
\caption{
Symmetry coordinates used in the
phonon calculations for four irreducible representations
of orthorhombic KNbO$_3$.
}
\label{tab:symmet}
\begin{tabular}{llll}
 \multicolumn{1}{c}{$A_1$} &
 \multicolumn{1}{c}{$A_2$} &
 \multicolumn{1}{c}{$B_1$} &
 \multicolumn{1}{c}{$B_2$} \\
\hline
 $S_1    = Z_1-\frac{1}{2}(Z_4+Z_5)$ &
 $S_5    = X_4-X_5$ &
 $S_6    = Y_1-\frac{1}{2}(Y_4+Y_5)$ &
 $S_{10} = X_1-\frac{1}{2}(X_4+X_5)$ \\
 $S_2    = Z_2-\frac{1}{2}(Z_4+Z_5)$ & &
 $S_7    = Y_2-\frac{1}{2}(Y_4+Y_5)$ &
 $S_{11} = X_2-\frac{1}{2}(X_4+X_5)$ \\
 $S_3    = Z_3-\frac{1}{2}(Z_4+Z_5)$ & &
 $S_8    = Y_3-\frac{1}{2}(Y_4+Y_5)$ &
 $S_{12} = X_3-\frac{1}{2}(X_4+X_5)$ \\
 $S_4    = Y_4-Y_5$ & &
 $S_9    = Z_4-Z_5$ &  \\
\end{tabular}
\end{table}

\begin{table}
\caption{
4th order polynomial fit of the total energy (in mRy)
in terms of the symmetry coordinates of
the $B_2$ representation (in a.u.).
}
\label{tab:polynom}
\begin{tabular}{cccdccccdccccd}
\multicolumn{3}{c}{powers of} & & &
\multicolumn{3}{c}{powers of} & & &
\multicolumn{3}{c}{powers of} & \\
{}~$S_{10}$~&~$S_{11}$~&~$S_{12}$~&
\multicolumn{1}{c}{\raisebox{2.0ex}[0pt]{Coef.}} &~~~~&
{}~$S_{10}$~&~$S_{11}$~&~$S_{12}$~&
\multicolumn{1}{c}{\raisebox{2.0ex}[0pt]{Coef.}} &~~~~&
{}~$S_{10}$~&~$S_{11}$~&~$S_{12}$~&
\multicolumn{1}{c}{\raisebox{2.0ex}[0pt]{Coef.}} \\
\hline
 2 & 0 & 0 &      30.70 & &
 3 & 1 & 0 &    $-$7.02 & &
 1 & 1 & 2 & $-$1316.91 \\
 1 & 1 & 0 &    $-$2.84 & &
 3 & 0 & 1 &  $-$300.44 & &
 1 & 0 & 3 &  $-$876.50 \\
 1 & 0 & 1 &   $-$61.27 & &
 2 & 2 & 0 &   $-$57.52 & &
 0 & 4 & 0 &     414.92 \\
 0 & 2 & 0 &    $-$2.76 & &
 2 & 1 & 1 &    1160.64 & &
 0 & 3 & 1 & $-$1359.44 \\
 0 & 1 & 1 &      57.91 & &
 2 & 0 & 2 &     847.09 & &
 0 & 2 & 2 &    1549.71 \\
 0 & 0 & 2 &      83.73 & &
 1 & 3 & 0 &      41.03 & &
 0 & 1 & 3 &  $-$925.13 \\
 4 & 0 & 0 &      84.82 & &
 1 & 2 & 1 &     386.80 & &
 0 & 0 & 4 &     214.28 \\
\end{tabular}
\end{table}

\begin{table}
\caption{
Calculated $\Gamma$-TO frequencies and eigenvectors
in orthorhombic KNbO$_3$.
}
\label{calc:knoort}
\begin{tabular}{lddddddd}
  &  Frequency & &
 \multicolumn{5}{c}{\rule[-2mm]{0mm}{6mm}Eigenvectors} \\
\cline{4-8}
 \raisebox{2.5ex}[0pt]{Symm.} &
 (cm$^{-1}$) &
 \raisebox{2.5ex}[0pt]{Polarization~~} &
 \rule[-2mm]{0mm}{6mm}
 K & Nb & O$_1$ & O$_2$ & O$_3$ \\
\hline
 \rule[0mm]{0mm}{3.5mm}
 &   &  $y$.&         &         &         &    0.07 & $-$0.07 \\
 & \raisebox{1.5ex}[0pt]{186} &
   \raisebox{1.5ex}[0pt]{$\Bigl\{$~~}
        $z$.& $-$0.86 &    0.48 & $-$0.03 &    0.10 &    0.10 \\
 &   &  $y$.&         &         &         &    0.34 & $-$0.34 \\
 & \raisebox{1.5ex}[0pt]{257} &
   \raisebox{1.5ex}[0pt]{$\Bigl\{$~~}
        $z$.&    0.15 & $-$0.00 & $-$0.77 &    0.28 &    0.28 \\
   \raisebox{2.0ex}[0pt]{$A_1$}
 &   &  $y$.&         &         &         &    0.25 & $-$0.25 \\
 & \raisebox{1.5ex}[0pt]{307} &
   \raisebox{1.5ex}[0pt]{$\Bigl\{$~~}
        $z$.& $-$0.14 & $-$0.48 &    0.50 &    0.44 &    0.44 \\
 &   &  $y$.&         &         &         &    0.57 & $-$0.57 \\
 \rule[-2mm]{0mm}{2mm}
 & \raisebox{1.5ex}[0pt]{593} &
   \raisebox{1.5ex}[0pt]{$\Bigl\{$~~}
        $z$.&    0.08 &    0.15 &    0.25 & $-$0.37 & $-$0.37 \\
\hline
 $A_2$ \rule[-2mm]{0mm}{5.5mm} &
 224 &  $x$.&         &         &         &    0.71 & $-$0.71 \\
\hline
 \rule[0mm]{0mm}{3.5mm}
 &   &  $y$.& $-$0.85 &    0.50 &    0.00 &    0.06 &    0.06 \\
 & \raisebox{1.5ex}[0pt]{146} &
   \raisebox{1.5ex}[0pt]{$\Bigl\{$~~}
        $z$.&         &         &         &    0.08 & $-$0.08 \\
 &   &  $y$.&    0.07 & $-$0.09 & $-$0.63 &    0.37 &    0.37 \\
 & \raisebox{1.5ex}[0pt]{232} &
   \raisebox{1.5ex}[0pt]{$\Bigl\{$~~}
        $z$.&         &         &         &    0.39 & $-$0.39 \\
   \raisebox{2.0ex}[0pt]{$B_1$}
 &   &  $y$.& $-$0.19 & $-$0.46 &    0.67 &    0.36 &    0.36 \\
 & \raisebox{1.5ex}[0pt]{297} &
   \raisebox{1.5ex}[0pt]{$\Bigl\{$~~}
        $z$.&         &         &         &    0.16 & $-$0.16 \\
 &   &  $y$.&    0.13 &    0.12 &    0.26 & $-$0.37 & $-$0.37 \\
 \rule[-2mm]{0mm}{2mm}
 & \raisebox{1.5ex}[0pt]{528} &
   \raisebox{1.5ex}[0pt]{$\Bigl\{$~~}
        $z$.&         &         &         &    0.56 & $-$0.56 \\
\hline
 \rule[ 0mm]{0mm}{3.5mm}
       & 82$i$ &
        $x$.&    0.48 & $-$0.69 &    0.43 &    0.24 &    0.24 \\
 $B_2$ &  185  &
        $x$.& $-$0.73 &    0.01 &    0.32 &    0.42 &    0.42 \\
 \rule[-2mm]{0mm}{2mm}
       &  467  &
        $x$.&    0.13 & $-$0.10 & $-$0.79 &    0.42 &    0.42 \\
\end{tabular}
\end{table}

\begin{table}
\caption{
Calculated and measured frequencies of $\Gamma$-TO phonons
in orthorhombic KNbO$_3$.
}
\label{exp:knoort}
\begin{tabular}{cccccc}
 & & Calculated \\
\raisebox{2.5ex}[0pt]{Mode}  &
\raisebox{2.5ex}[0pt]{Symm.} &
 frequency (cm$^{-1}$) &
\raisebox{2.5ex}[0pt]{Expt.\tablenotemark[1]}  &
\raisebox{2.5ex}[0pt]{Expt.\tablenotemark[2]}  &
\raisebox{2.5ex}[0pt]{Expt.\tablenotemark[3]}  \\
\hline
        & $B_2$ & soft &  56 &  40   &  59   \\
 TO$_1$ & $B_1$ &  232 & 243 & 249   &       \\
        & $A_1$ &  257 & 290 & 281.5 &       \\
\\
        & $B_2$ &  185 & 195 & 196.5 & 197.5 \\
 TO$_2$ & $B_1$ &  146 & 187 & 192   &       \\
        & $A_1$ &  186 & 190 & 193   &       \\
\\
        & $B_2$ &  467 & 511 & 513   & 516   \\
 TO$_3$ & $B_1$ &  528 & 534 & 534   &       \\
        & $A_1$ &  593 & 607 & 606.5 &       \\
\\
        & $B_1$ &  297 & 270 & 270   &       \\
 TO$_4$ & $A_1$ &  307 & 299 & 297   &       \\
        & $A_2$ &  224 & 283 & 283   &       \\
\end{tabular}
\tablenotetext[1]{Raman spectroscopy; Ref. \cite{bh76}.}
\tablenotetext[2]{Raman spectroscopy; Ref. \cite{qbkr76}.}
\tablenotetext[3]{Infrared spectroscopy; Ref. \cite{fmsg84}.}
\end{table}

\setlength{\unitlength}{0.65cm}
%
%
\begin{picture}(  30.000,  20.000)
\thinlines
 \put(11,10){    
 \put(-13.,0.){\vector(1,0){26.}}
 \put(-10., 0.){\line(0,-1){0.2}}
 \put(-10.,-1.2){\makebox(0,0){\Large $-$0.1}}
 \put( -5., 0.){\line(0,-1){0.2}}
 \put( -5.,-1.2){\makebox(0,0){\Large $-$0.05}}
 \put(  5., 0.){\line(0,-1){0.2}}
 \put(  5.,-1.2){\makebox(0,0){\Large 0.05}}
 \put( 10., 0.){\line(0,-1){0.2}}
 \put( 10.,-1.2){\makebox(0,0){\Large 0.1}}
 \put( 13.,-1.2){\makebox(0,0){\Large $\Delta_z$ (\AA)}}

 \put(  0.,-8.){\vector(0,1){16.}}
 \put(  0.,-5.){\line(1, 0){0.2}}
 \put(  1.5,-5.){\makebox(0,0){\Large $-$0.05}}
 \put(  0., 5.){\line(1, 0){0.2}}
 \put(  1.5, 5.){\makebox(0,0){\Large 0.05}}
 \put(  2., 8.){\makebox(0,0){\Large $\Delta_x$ (\AA)}}

 \put( 11.50, 2.58){\circle*{1.3}}
 \put( 13.00, 2.58){\makebox(0,0){\Large K}}
 \put( -9.49,-2.32){\circle*{1.3}}
 \put( -7.99,-2.32){\makebox(0,0){\Large Nb}}
 \put(  9.15, 3.27){\circle*{1.3}}
 \put(  7.65, 3.27){\makebox(0,0){\Large O(I)}}
 \put(  8.99, 1.98){\circle*{1.3}}
 \put(  7.49, 1.98){\makebox(0,0){\Large O(II)}}
\thicklines
 \put(  1.32, 0.79){\circle{1.3}}
 \put(  2.82, 0.79){\makebox(0,0){\Large K}}
 \put( -6.07,-4.43){\circle{1.3}}
 \put( -4.57,-4.43){\makebox(0,0){\Large Nb}}
 \put( 10.69, 7.94){\circle{1.3}}
 \put(  9.19, 7.94){\makebox(0,0){\Large O}}
}
\put(16.0, -12.){\Huge Fig. 1}
\end{picture}
\newpage
\thispagestyle{empty}
\setlength{\unitlength}{0.24cm}
%
%
\begin{picture}(  50.000,  75.000)
 \put(0,25){    
 \thicklines
 \put(  28.3,    32.500){\line(0,-1){8.}}
 \put(  28.3,    15.500){\vector(0,-1){8.}}
 \put(  28.3,    34.500){\makebox(0,0){\Large center of cube}}
 \put(   2.273,    .000){\vector(1,0){50.0}}
 \put(  11.364,   -.500){\line(0,1){0.5}}
 \put(  20.455,   -.500){\line(0,1){0.5}}
 \put(  29.545,   -.500){\line(0,1){0.5}}
 \put(  38.636,   -.500){\line(0,1){0.5}}
 \put(  47.727,   -.500){\line(0,1){0.5}}
 \put(  11.364,  -2.500){\makebox(0,0){\huge 0}}
 \put(  29.545,  -2.500){\makebox(0,0){\huge 0.4}}
 \put(  47.727,  -2.500){\makebox(0,0){\huge 0.8}}
 \put(  61.000,  -2.800){\makebox(0,0){\huge $\Delta_z$ (a.u.)}}
 \put(   2.273,    .000){\vector(0,1){50.0}}
 \put(   1.773,   7.143){\line(1,0){0.5}}
 \put(   1.773,  14.286){\line(1,0){0.5}}
 \put(   1.773,  21.429){\line(1,0){0.5}}
 \put(   1.773,  28.571){\line(1,0){0.5}}
 \put(   1.773,  35.714){\line(1,0){0.5}}
 \put(  -0.273,   7.143){\makebox(0,0){\huge 0}}
 \put(  -0.273,  21.429){\makebox(0,0){\huge 10}}
 \put(  -0.273,  35.714){\makebox(0,0){\huge 20}}
 \put(  -3.222,  47.000){\makebox(0,0){\huge $\Delta E$}}
 \put(  -3.222,  43.000){\makebox(0,0){\huge (mRy)}}
\newsavebox{\markname}
\savebox{\markname}(0,0)[bl]{
 \put(0,0){\circle{1.5}}  
}
 \put(   4.545,   9.629){\usebox{\markname}}
 \put(  11.364,   7.143){\usebox{\markname}}
 \put(  18.182,   9.243){\usebox{\markname}}
 \put(  28.636,  18.957){\usebox{\markname}}
 \put(  45.455,  48.357){\usebox{\markname}}
 \put(  35., 15.){\usebox{\markname}}
 \put(  40., 15.){\makebox(0,0){\huge a}}
 \put(  2.27273, 11.73901){\line( 1,-1){1.53}}
 \put(  3.79805, 10.24383){\line( 4,-3){1.53}}
 \put(  5.32336,  9.06661){\line( 5,-3){1.53}}
 \put(  6.84869,  8.18676){\line( 5,-2){1.53}}
 \put(  8.37401,  7.58486){\line( 4,-1){1.53}}
 \put(  9.89933,  7.14272){\line( 1, 0){1.53}}
 \put( 11.42465,  7.17095){\line( 1, 0){1.53}}
 \put( 12.94997,  7.27095){\line( 4, 1){1.53}}
 \put( 14.47529,  7.61098){\line( 3, 1){1.53}}
 \put( 16.00059,  8.15005){\line( 2, 1){1.53}}
 \put( 17.52591,  8.87600){\line( 5, 3){1.53}}
 \put( 19.05123,  9.77790){\line( 3, 2){1.53}}
 \put( 20.57659, 10.84600){\line( 5, 4){1.53}}
 \put( 22.10191, 12.07174){\line( 1, 1){1.53}}
 \put( 23.62723, 13.64784){\line( 1, 1){1.53}}
 \put( 25.15255, 15.16814){\line( 1, 1){1.53}}
 \put( 26.67786, 16.62776){\line( 5, 6){1.53}}
 \put( 28.20318, 18.42297){\line( 4, 5){1.53}}
 \put( 29.72850, 20.35129){\line( 3, 4){1.53}}
 \put( 31.25382, 22.41143){\line( 2, 3){1.53}}
 \put( 32.77914, 24.60329){\line( 2, 3){1.53}}
 \put( 34.30445, 26.92800){\line( 3, 5){1.53}}
 \put( 35.82982, 29.38800){\line( 3, 5){1.53}}
 \put( 37.35514, 31.98671){\line( 1, 2){1.53}}
 \put( 38.88045, 34.72886){\line( 1, 2){1.53}}
 \put( 40.40577, 37.62057){\line( 1, 2){1.53}}
 \put( 41.93109, 40.66871){\line( 1, 2){1.53}}
 \put( 43.45641, 43.88200){\line( 1, 2){1.53}}
 \put( 44.98173, 47.26986){\line( 2, 5){1.53}}
\newsavebox{\marknamb}
\savebox{\marknamb}(0,0)[bl]{
 \put(0,0){\circle*{1.5}}  
}
 \put(   2.727,  11.071){\usebox{\marknamb}}
 \put(  11.364,   7.143){\usebox{\marknamb}}
 \put(  28.636,  18.629){\usebox{\marknamb}}
 \put(  45.455,  46.686){\usebox{\marknamb}}
 \put(  35., 10.){\usebox{\marknamb}}
 \put(  40., 10.){\makebox(0,0){\huge b}}
 \put(  2.27273, 11.48067){\line( 6,-5){1.53}}
 \put(  3.79805, 10.18937){\line( 3,-2){1.53}}
 \put(  5.32336,  9.12993){\line( 2,-1){1.53}}
 \put(  6.84869,  8.29907){\line( 5,-2){1.53}}
 \put(  8.37401,  7.69317){\line( 4,-1){1.53}}
 \put(  9.89933,  7.19830){\line( 1, 0){1.53}}
 \put( 11.42465,  7.08444){\line( 1, 0){1.53}}
 \put( 12.94997,  7.18444){\line( 6, 1){1.53}}
 \put( 14.47529,  7.43604){\line( 3, 1){1.53}}
 \put( 16.00059,  7.88986){\line( 5, 2){1.53}}
 \put( 17.52591,  8.54044){\line( 2, 1){1.53}}
 \put( 19.05123,  9.38197){\line( 3, 2){1.53}}
 \put( 20.57659, 10.40836){\line( 5, 4){1.53}}
 \put( 22.10191, 11.61317){\line( 1, 1){1.53}}
 \put( 23.62723, 13.18969){\line( 1, 1){1.53}}
 \put( 25.15255, 14.73089){\line( 1, 1){1.53}}
 \put( 26.67786, 16.22939){\line( 5, 6){1.53}}
 \put( 28.20318, 18.07753){\line( 3, 4){1.53}}
 \put( 29.72850, 20.06734){\line( 3, 4){1.53}}
 \put( 31.25382, 22.19057){\line( 2, 3){1.53}}
 \put( 32.77914, 24.43857){\line( 2, 3){1.53}}
 \put( 34.30445, 26.80243){\line( 3, 5){1.53}}
 \put( 35.82982, 29.27286){\line( 3, 5){1.53}}
 \put( 37.35514, 31.84057){\line( 3, 5){1.53}}
 \put( 38.88045, 34.49543){\line( 1, 2){1.53}}
 \put( 40.40577, 37.22743){\line( 1, 2){1.53}}
 \put( 41.93109, 40.02600){\line( 1, 2){1.53}}
 \put( 43.45641, 42.88043){\line( 1, 2){1.53}}
 \put( 44.98173, 45.77971){\line( 1, 2){1.53}}
 }
\newsavebox{\stleft}
\savebox{\stleft}(0,0)[bl]{
 \put( 0, 0){\line(-1,0){1.0}}
 \put(-1.,0){\circle*{0.8}}
 \thicklines
 \put( 0.,-0.2){\line(0,1){0.4}}
 \thinlines
}
\newsavebox{\stright}
\savebox{\stright}(0,0)[bl]{
 \put( 0, 0){\line(1,0){1.0}}
 \put( 1.,0){\circle*{0.8}}
 \thicklines
 \put( 0.,-0.2){\line(0,1){0.4}}
 \thinlines
}
\newsavebox{\biglayer}
\savebox{\biglayer}(0,0)[bl]{
 \put(-14., 0.){\line(3, 1){12.}}
 \put(-10.,-1.){\line(3, 1){12.}}
 \put( -6.,-2.){\line(3, 1){12.}}
 \put( -2.,-3.){\line(3, 1){12.}}
 \put(  2.,-4.){\line(3, 1){12.}}
 \put(-14., 0.){\line(4,-1){16.}}
 \put(-11., 1.){\line(4,-1){16.}}
 \put( -8., 2.){\line(4,-1){16.}}
 \put( -5., 3.){\line(4,-1){16.}}
 \put( -2., 4.){\line(4,-1){16.}}
}

\put(15., 8.){  
 \multiput(0.,0.)(0.,-8.){2}{
  \put(0.,0.){\usebox{\biglayer}}
  \multiput(-10.5, 0.)( 6., 2.){2}{\usebox{\stleft}}
  \multiput( -7.5, 1.)( 6., 2.){2}{\usebox{\stright}}
  \multiput( -2.5,-2.)( 6., 2.){2}{\usebox{\stleft}}
  \multiput(  0.5,-1.)( 6., 2.){2}{\usebox{\stright}}
  \multiput( -6.5,-1.)( 3., 1.){4}{\usebox{\stright}}
  \multiput(  1.5,-3.)( 3., 1.){4}{\usebox{\stright}}
 }
 \thicklines
 \put(-14.3,  0.){\line( 3, 1){6.3}}
 \put(-14.3,  0.){\line( 4,-1){8.3}}
 \put(  0.3,  0.){\line(-4, 1){8.3}}
 \put(  0.3,  0.){\line(-3,-1){6.3}}
 \thinlines
 \put(0.,-18.){\huge a}
}

\put(46., 8.){  
 \multiput(0.,0.)(0.,-8.){2}{
  \put(0.,0.){\usebox{\biglayer}}
  \multiput(-10.5, 0.)( 7., 0.){4}{\usebox{\stleft}}
  \multiput( -7.5, 1.)( 7., 0.){3}{\usebox{\stright}}
  \multiput( -6.5,-1.)( 7., 0.){3}{\usebox{\stright}}
  \multiput( -4.5, 2.)( 7., 0.){2}{\usebox{\stright}}
  \multiput( -2.5,-2.)( 7., 0.){2}{\usebox{\stright}}
  \put( -1.5, 3.){\usebox{\stright}}
  \put(  1.5,-3.){\usebox{\stright}}
 }
 \thicklines
 \put(-14.3,  0.){\line( 3, 1){12.3}}
 \put(-14.3,  0.){\line( 4,-1){ 4.3}}
 \put(  2.3,  3.){\line(-4, 1){ 4.3}}
 \put(  2.3,  3.){\line(-3,-1){12.3}}
 \thinlines
 \put(0.,-18.){\huge b}
}
\put(52.,-25.){\Huge Fig. 2}
\end{picture}
\newpage
\thispagestyle{empty}
\setlength{\unitlength}{0.30cm}
%
%
\begin{picture}(  40.000,  70.000)
 \put(4,23){    
 \thicklines
 \put(    .000,    .000){\vector(1,0){45.0}}
 \put(  13.333,   -.500){\line(0,1){0.5}}
 \put(  26.667,   -.500){\line(0,1){0.5}}
 \put(  40.000,   -.500){\line(0,1){0.5}}
 \put(   0.000,  -2.000){\makebox(0,0){\Huge 0}}
 \put(  13.333,  -2.000){\makebox(0,0){\Huge 0.05}}
 \put(  26.667,  -2.000){\makebox(0,0){\Huge 0.10}}
 \put(  43.000,  -2.500){\makebox(0,0){\Huge $\Delta_x$ (a.u.)}}
 \put(    .000,    .000){\vector(0,1){43.0}}
 \put(   -.500,   6.667){\line(1,0){0.5}}
 \put(   -.500,  13.333){\line(1,0){0.5}}
 \put(  -2.500,  13.333){\makebox(0,0){\Huge 0}}
 \put(   -.500,  20.000){\line(1,0){0.5}}
 \put(   -.500,  26.667){\line(1,0){0.5}}
 \put(  -2.500,  26.667){\makebox(0,0){\Huge 0.1}}
 \put(   -.500,  33.333){\line(1,0){0.5}}
 \put(   -.500,  40.000){\line(1,0){0.5}}
 \put(  -2.500,  40.000){\makebox(0,0){\Huge 0.2}}
 \put(    .000,  51.000){\makebox(0,0){\Huge $\Delta E$}}
 \put(    .000,  48.000){\makebox(0,0){\Huge (mRy)}}

 \put(  38.500,  22.000){\makebox(0,0){\Huge a}}
 \put(  34.000,  10.000){\makebox(0,0){\Huge b}}
 \put(   7.000,  30.000){\makebox(0,0){\Huge c}}

\newsavebox{\marknamk}
\savebox{\marknamk}(0,0)[bl]{
 \put(0,0){\circle{1.0}}  
}
 \put(    .000,  13.333){\usebox{\marknamk}}
 \put(   6.000,  12.000){\usebox{\marknamk}}
 \put(  12.000,  10.667){\usebox{\marknamk}}
 \put(  18.000,  10.667){\usebox{\marknamk}}
 \put(  24.027,  12.000){\usebox{\marknamk}}
 \put(  30.027,  16.000){\usebox{\marknamk}}
 \put(  36.027,  22.667){\usebox{\marknamk}}
 \put(   .26848, 12.92414){\line( 1, 0){2.15}}
 \put(  2.41613, 12.92414){\line( 6,-1){2.15}}
 \put(  4.56379, 12.49161){\line( 5,-1){2.15}}
 \put(  6.71144, 12.04577){\line( 4,-1){2.15}}
 \put(  8.85909, 11.52772){\line( 4,-1){2.15}}
 \put( 11.00672, 11.01791){\line( 5,-1){2.15}}
 \put( 13.15437, 10.60195){\line( 1, 0){2.15}}
 \put( 15.30203, 10.56059){\line( 1, 0){2.15}}
 \put( 17.44968, 10.55985){\line( 1, 0){2.15}}
 \put( 19.59733, 10.64308){\line( 4, 1){2.15}}
 \put( 21.74499, 11.22624){\line( 5, 2){2.15}}
 \put( 23.89264, 12.09874){\line( 2, 1){2.15}}
 \put( 26.04029, 13.23180){\line( 3, 2){2.15}}
 \put( 28.18800, 14.59696){\line( 4, 3){2.15}}
 \put( 30.33547, 16.19739){\line( 1, 1){2.15}}
 \put( 32.48320, 18.35495){\line( 1, 1){2.15}}
 \put( 34.63093, 20.57641){\line( 3, 5){2.15}}
\newsavebox{\marknami}
\savebox{\marknami}(0,0)[bl]{
 \put(0,0){\circle*{1.0}}  
}
 \put(    .000,  13.333){\usebox{\marknami}}
 \put(   6.000,  12.000){\usebox{\marknami}}
 \put(  12.000,  10.667){\usebox{\marknami}}
 \put(  18.000,   8.000){\usebox{\marknami}}
 \put(  24.027,   6.667){\usebox{\marknami}}
 \put(  30.027,   9.333){\usebox{\marknami}}
 \put(  36.027,  32.000){\usebox{\marknami}}
 \put(   .26848, 13.04151){\line( 1, 0){2.15}}
 \put(  2.41613, 13.04151){\line( 6,-1){2.15}}
 \put(  4.56379, 12.59912){\line( 4,-1){2.15}}
 \put(  6.71144, 12.11251){\line( 3,-1){2.15}}
 \put(  8.85909, 11.48844){\line( 3,-1){2.15}}
 \put( 11.00672, 10.76445){\line( 3,-1){2.15}}
 \put( 13.15437,  9.97919){\line( 5,-2){2.15}}
 \put( 15.30203,  9.16947){\line( 3,-1){2.15}}
 \put( 17.44968,  8.37120){\line( 3,-1){2.15}}
 \put( 19.59733,  7.72633){\line( 3,-1){2.15}}
 \put( 21.74499,  6.99888){\line( 5,-1){2.15}}
 \put( 23.89264,  6.40345){\line( 1, 0){2.15}}
 \put( 26.04029,  6.45009){\line( 5, 2){2.15}}
 \put( 28.18800,  7.30964){\line( 1, 1){2.15}}
 \put( 30.33547,  9.50432){\line( 2, 5){2.15}}
 \put( 32.48320, 14.72889){\line( 1, 4){2.15}}
 \put( 34.63093, 23.20768){\line( 1, 6){2.15}}
\newsavebox{\marknamj}
\savebox{\marknamj}(0,0)[bl]{
 \put(0,0){\circle*{1.0}}  
}
 \put(    .000,  13.333){\usebox{\marknamj}}
 \put(   6.000,  40.000){\usebox{\marknamj}}
 \put(   .14317, 13.34851){\line( 1, 1){1.18}}
 \put(  1.28858, 14.56265){\line( 1, 3){1.15}}
 \put(  2.43400, 18.01979){\line( 1, 4){1.15}}
 \put(  3.57941, 22.82075){\line( 1, 6){2.80}}
 }
\thinlines

\newsavebox{\stup}
\savebox{\stup}(0,0)[bl]{
 \put( 0.,0. ){\line(-1,0){1.0}}
 \put(-1.,0. ){\line( 0,1){0.4}}
 \put(-1.,0.4){\circle*{0.8}}
 \thicklines
 \put( 0.,-0.2){\line(0,1){0.4}}
 \thinlines
}
\newsavebox{\stdn}
\savebox{\stdn}(0,0)[bl]{
 \put( 0., 0. ){\line(-1, 0){1.0}}
 \put(-1., 0. ){\line( 0,-1){0.4}}
 \put(-1.,-0.4){\circle*{0.8}}
 \thicklines
 \put( 0.,-0.2){\line(0,1){0.4}}
 \thinlines
}
\newsavebox{\smalllayer}
\savebox{\smalllayer}(0,0)[bl]{
 \put(-7., 0.){\line(3, 1){6.}}
 \put(-3.,-1.){\line(3, 1){6.}}
 \put( 1.,-2.){\line(3, 1){6.}}
 \put(-7., 0.){\line(4,-1){8.}}
 \put(-4., 1.){\line(4,-1){8.}}
 \put(-1., 2.){\line(4,-1){8.}}
}
\put( 8.,11.){  
 \multiput(0.,0.)(0.,-4.){3}{
  \put(0.,0.){\usebox{\smalllayer}}
  \multiput( -3.5, 0.)( 3., 1.){2}{\usebox{\stup}}
  \multiput(  0.5,-1.)( 3., 1.){2}{\usebox{\stup}}
 }
 \thicklines
 \put( -7.1,  0. ){\line(3, 1){ 3.1}}
 \put( -7.1,  0. ){\line(4,-1){ 4.1}}
 \put( -3.0, -1.1){\line(3, 1){ 3.1}}
 \put( -4.0,  1.1){\line(4,-1){ 4.1}}
 \thinlines
 \put( 0.,-15.){\Huge a}
}
\put(25.,11.){  
 \multiput(0.,0.)(0.,-4.){3}{
  \put(0.,0.){\usebox{\smalllayer}}
  \put( -3.5, 0.){\usebox{\stup}}
  \put(  3.5, 0.){\usebox{\stup}}
  \put( -0.5, 1.){\usebox{\stdn}}
  \put(  0.5,-1.){\usebox{\stdn}}
 }
 \thicklines
 \put( -7.1,  0. ){\line(3, 1){ 6.2}}
 \put( -7.1,  0. ){\line(4,-1){ 4.1}}
 \put( -3.0, -1.1){\line(3, 1){ 6.2}}
 \put( -1.0,  2.1){\line(4,-1){ 4.1}}
 \thinlines
 \put( 0.,-15.){\Huge b}
}
\put(42.,11.){  
 \multiput(0.,0.)(0.,-8.){2}{
  \put(0.,0.){\usebox{\smalllayer}}
  \put( -3.5, 0.){\usebox{\stup}}
  \put(  3.5, 0.){\usebox{\stup}}
  \put( -0.5, 1.){\usebox{\stdn}}
  \put(  0.5,-1.){\usebox{\stdn}}
 }
 \put(0.,-4.){
  \put(0.,0.){\usebox{\smalllayer}}
  \put( -3.5, 0.){\usebox{\stdn}}
  \put(  3.5, 0.){\usebox{\stdn}}
  \put( -0.5, 1.){\usebox{\stup}}
  \put(  0.5,-1.){\usebox{\stup}}
 }
 \thicklines
 \put( -7.1,  0. ){\line(3, 1){ 6.2}}
 \put( -7.1,  0. ){\line(4,-1){ 4.1}}
 \put( -3.0, -1.1){\line(3, 1){ 6.2}}
 \put( -1.0,  2.1){\line(4,-1){ 4.1}}
 \thinlines
 \put( 0.,-15.){\Huge c}
}
\put(44.,-11.){\Huge Fig. 3}
\end{picture}

\begin{references}
\bibitem{note1}  Of course the accuracy of 0.01 mRy applies
                 only to the trends in the total energy
                 calculated for different geometries but
                 with otherwise identical setup,
                 and not to the absolute total energy values
                 which are sensitive, e.g., to the choice
                 of muffin-tin sphere radii within much
                 broader limits.
\bibitem{note2}  This is not in contradiction with the
		 eight-site model the essential statement of which
		 is that Nb atoms occupy off-center
		 [111]-type displaced positions in all
		 cristallographic phases of KNbO$_3$.
		 Our calculation result just cited indicates
		 that hoppings over such displaced positions
		 do not occur independently in adjacent
		 unit cells, but rather exhibit some
		 spatial coherence.
\end{references}
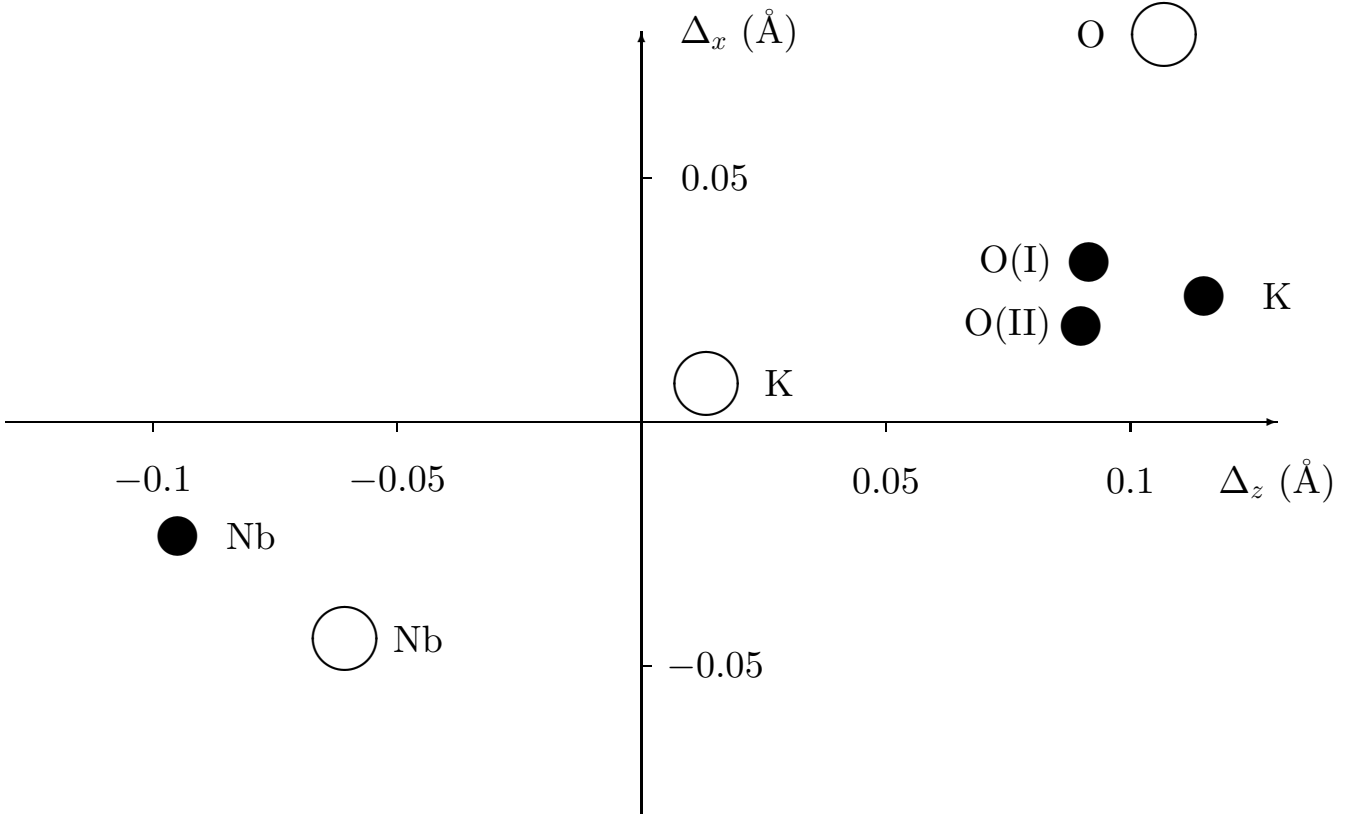
\end{document}